\documentclass[sigchi]{acmart}

\AtBeginDocument{%
  \providecommand\BibTeX{{%
    \normalfont B\kern-0.5em{\scshape i\kern-0.25em b}\kern-0.8em\TeX}}}

\copyrightyear{2019}
\acmYear{2019}
\setcopyright{acmlicensed}

\usepackage{tikz}

\usepackage{marvosym}

\usepackage{pgfplots}
\pgfplotsset{compat=1.8}
\usepgfplotslibrary{statistics}
\makeatletter
\pgfplotsset{
    boxplot prepared from table/.code={
        \def\tikz@plot@handler{\pgfplotsplothandlerboxplotprepared}%
        \pgfplotsset{
            /pgfplots/boxplot prepared from table/.cd,
            #1,
        }
    },
    /pgfplots/boxplot prepared from table/.cd,
        table/.code={\pgfplotstablecopy{#1}\to\boxplot@datatable},
        row/.initial=0,
        make style readable from table/.style={
            #1/.code={
                \pgfplotstablegetelem{\pgfkeysvalueof{/pgfplots/boxplot prepared from table/row}}{##1}\of\boxplot@datatable
                \pgfplotsset{boxplot/#1/.expand once={\pgfplotsretval}}
            }
        },
        make style readable from table=lower whisker,
        make style readable from table=upper whisker,
        make style readable from table=lower quartile,
        make style readable from table=upper quartile,
        make style readable from table=median,
        make style readable from table=lower notch,
        make style readable from table=upper notch
}
\makeatother

\pgfplotstableread{
    lw lq med  uq uw name
     1.571429  2.964286 3.428571  4 4.85713 FFbCF
     2  2.857143 3.335255    3.857143 4.571429  WFFbR
     2  3 3.441691  3.857143 4.714286 FFbR
     1.285714  2.285714 2.804665  3.321429 4.428571 Baseline
}\datatable

\begin{document}

%
\title{ Personalized, Health-Aware Recipe Recommendation: An Ensemble Topic Modeling Based Approach }

%



\author{Mansura A. Khan}
\authornote{
Insight Centre for Data Analytics, University College Dublin}
\email{mansura.khan@ucd.ie}

\affiliation{%
  \institution{}
  \streetaddress{}
  \city{}
  \state{}
  \postcode{Dublin-4}
}

\author{Ellen Rushe}
\authornotemark[1]
\email{ellen.rushe@ucdconnect.ie}

\affiliation{%
  \institution{}
  \streetaddress{}
  \city{}
  \state{}
  \postcode{Dublin-4}
}

\author{Barry Smyth}
\authornotemark[1]
\email{barry.smyth@ucd.ie}
\affiliation{%
 \institution{ }
  \streetaddress{}
  \city{}
  \country{}
}
\author{David Coyle}
\authornotemark[1]
\email{d.coyle@ucd.ie}
\affiliation{%
  \institution{ }
  \streetaddress{}
  \city{}
  \country{}}

%
\renewcommand{\shortauthors}{Mansura A Khan et al.}

%
\begin{abstract}
Food choices are personal and complex and have a significant impact on our long-term health and quality of life. By helping users to make informed and satisfying decisions, Recommender Systems (RS) have the potential to support users in making healthier food choices. Intelligent users-modeling is a key challenge in achieving this potential. This paper investigates Ensemble Topic Modelling (\(E_{ns}TM\)) based \textbf{Feature Identification} techniques for efficient user-modeling and recipe recommendation. It builds on findings in \(E_{ns}TM\) to propose a reduced data representation format and a smart user-modeling strategy that makes capturing user-preference fast, efficient and interactive. This approach enables personalization, even in a cold-start scenario. This paper proposes two different \(E_{ns}TM\) based and one Hybrid \(E_{ns}TM\) based recommenders. We compared all three \(E_{ns}TM\) based variations through a user study with 48 participants, using a large-scale, real-world corpus of 230,876 recipes, and compare against a conventional Content Based (CB) approach. \(E_{ns}TM\) based recommenders performed significantly better than the CB approach. Besides acknowledging multi-domain contents such as taste, demographics and cost, our proposed approach also considers user's nutritional preference and assists them finding recipes under diverse nutritional categories. Furthermore, it provides excellent coverage and enables implicit understanding of user's food practices. Subsequent analysis also exposed correlation between certain features and a healthier lifestyle.


\end{abstract}



%

%
\maketitle\vspace{-1.0em}
\section{Introduction}
Food has a direct, complex and multifaceted relationship with our lifestyle and personality. People have preferences regarding activities around food, such as cooking, plating, grocery and eating-out. Studies showed people are becoming more mindful towards healthier lifestyles and the fact that healthy eating/cooking impacts psychosocial and physical well-being  \cite{article:HowPeopleInterpretHealthyEating}. However, finding food-ideas/recipes that acknowledge one's circumstance and preference remains a challenge for many people. Food Recommender Systems (FRS) have the potential to assist users in navigating through the overwhelming amount of online resources on food/recipes and guide them towards healthier choices. 
World Health Organization figures suggest that 1.9 billion adults and 41 million children under the age of 5 are clinically overweight and more than 691 million people are obese \cite{website:WHO-fact-sheet}. FRS have the potential to become an important technology for addressing the global crisis of obesity and malnutrition by helping to make diet and nutritional guidelines available to mass population. This is well-reflected in the dramatic uplift of research interest on FRS in recent years \cite{DBLP:journals/corr/abs-1711-02760}.

Recommending food is challenging as our choices are defined by many cross-domain factors including demographic and contextual factors, health awareness, social and ethical factors, together with practical considerations such as cost, cooking time and methods, and the availability of ingredients. In order to develop effective FRS, we must design user models that capture user data across these diverse factors. Approaches are also required that enable Recommender Systems  (RS) to fit user's preference data on a massive information space around food. As Teng et al. note, there are millions of food-items/recipes as different ingredients are grown at different geographical locations and recipes originate from different cultural groups worldwide \cite{Teng:2012:RRU:2380718:2380757}. In this context coverage and diversity are important constraints, where coverage corresponds to the percentage of items for which a RS is able to generate a prediction \cite{Ge:2015:HFR:2792838.2796554}. Higher coverage enables the RS to implement varying diversity approaches and draw from more options. Taken together, these challenges necessitate FRS that can (1) identify the attributes/features which are significant for human food-choices, (2) capture user's preference on the identified features, (3) filter a large information-space, (4) generate recommendations efficiently and finally (5) guide users towards healthier choices. \vspace{0mm}

We explored Ensemble Topic Modelling (\(E_{ns}TM\)) \cite{MarkBelford} accompanied by a series of custom text-prepossessing to extract significant food features. The aim was to identify representative or agent contents of diverse domains connected to human food choice. In our study 288 features and their corresponding significance scores were extracted from a  corpus of 230,876 recipes.
As summarized in Table 1, the identified feature set is rich in contents representing multiple domains. The paper describes a foreshortened data representation format based on the extracted features which aims to reduce computational complexity of food recommendation.

We implemented three distinct \(E_{ns}TM\) based personalized FRS: a Food Feature based Recommender (FFbR), a Weighted Food Feature based Recommender (WFFbR), and a Food Feature based Collaborative Filtering (FFbCF). 
To evaluate these approaches we conducted a user study comparing \(E_{ns}TM\) based recommenders to a conventional Content Based (CB) approach. Results show that all \(E_{ns}TM\) based approaches significantly outperformed CB approach. In contrast to prior work, the \(E_{ns}TM\) approach also effectively supported recommendations across diverse social and cultural groups, even in a first recommendation scenario. Finally, the strong adaptation of the concept of dislike across all three methods proved effective in identifying user's food practice (e.g. vegetarian) and filtering accordingly. Further exploratory analysis exposed previously unknown pattern in user's interactions towards certain features.   
The existing correlation between healthier user-groups and certain food features argue for further research on feature based FRS with healthiness cues.

\vspace{-1em}
\section{RELATED WORK}
Previous research has produced seminal contributions towards FRS, aimed at ensuring user-preference, diversity and nutritional development in diet. Freyne et al. \cite{Freyne:2010:RFR:2149528.2149568,Freyne:2010:IFP:1719970.1720021} describe an ingredient-based approach where they inferred user's preference on a new recipe as the cumulative sum of his/her preference for each ingredient in that recipe.
This formed the basis of their novel user-based K-NN Collaborative Filtering (CF) approach \cite{Freyne:2010:IFP:1719970.1720021} which has been influential and was applied by others including \cite{ Harvey:2013:YYE:2651320.2651339,Sobecki:2006:AHR:2165946.2166064}. Subsequently, more advanced methods emerged for tackling different challenges such as, Teng et al. \cite{Teng:2012:RRU:2380718:2380757} used item-centric CF and applied an ingredient-network to identify similar recipes, where the ingredient-network was generated based on co-occurrence of ingredients within recipes and menus. Kuo et al. \cite{Kuo:2012:IMP:2390776.2390778} proposed a weighted graph based menu planning approach  where ingredients were grouped into subsets and each subset was considered as contents. However, while these approaches are very interesting, they focus purely on ingredients.

Ge et al. \cite{Ge:2015:UTL:2750511.2750528} proposed a method that leverages tags and latent factors to recommend recipes. Pinxteren et al. adopted a different approach \cite{vanPinxteren:2011:DRS:1943403.1943422} where, first they added custom annotations to each recipe in their corpus, then asked users to rate individual recipes and finally recommended recipes that share annotations with those rated positively by the user. This method was successful in addressing more food-choice factors, but the annotation set was relatively small and specific to their recipe corpus. As they mentioned, this limited  their FRS from automatically adopting to new user groups. Further notable work includes: Gu et al. \cite{Gu:2009:CFB:1701835.1701851} case-based FRS based on user's previous consumption cases; Sobeck et al. \cite{Sobecki:2006:AHR:2165946.2166064} hybrid FRS incorporating fuzzy inference with stereotype demographic filtering; and Bianca et al. \cite{Bobadilla:2012:CFA:2107778.2107834} hybrid model incorporating meta-heuristic and genetic algorithms. Elsweile et al.\cite{Elsweiler:2015:TAM:2792838.2799665} and Ueta et al. \cite{ Ueta:2011:RRS:2186633.2186642} discussed automatic meal planning approach to support balanced nutrition. While effective in constrained contexts, each of these approaches depends on sufficient pre-existing user preference data. They are thus susceptible to failure in cold-start scenarios \cite{Bobadilla:2012:CFA:2107778.2107834}. Trattner et al. \cite{Trattner2018} proposing a novel method to recommend recipes to people in a cold-start scenario.\vspace{0mm}

There was also a significant number of interesting research work producing domain specific knowledge to facilitate future research interests.\cite{ DBLP:journals/corr/abs-1711-02760} is a seminal work form Trattner et al. on summarizing, "to which extent current recommendation algorithms can adopt healthy recipes recommendation?" and "what resources are out there?". \cite{Rokicki:2016:PPG:2930238.2930248, Rokicki2018TheIO, Trattner:2017:RCI:3099023.3099072, Trattner:IHI:3038912:3052573} showed how online recipe repositories could be potential sources for knowledge discovery to support personalized and group-based recipe recommendations.
\cite{Elsweiler:EFC:3077136:3080826,TRATTNER:2019654, Harvey:2013:YYE:2651320.2651339} looked into patterns in users' online activity around food.\vspace{-1em}

\section{RECOMMENDER STRATEGIES}
To create a recipe data-set, we developed a web-scraper for geniuskitchen.com \cite{website:geniuskitchen}. Our final data-set comprises of 230,876 recipes. Each recipe was stored as a plain-text document that included information on ingredients, instructions, servings, cuisine, cooking-time, cooking-approach, cooking equipment, context, taste (e.g. sour or spicy) and nutrition data.

The first aim of our work was to uncover common food-features across the recipe data-set that could then be used to model user-preference and resolve user-to-recipe relationships. One traditional approach to achieving this is to apply TF-IDF \cite{articleTF_IDF}. This provides a term (word) frequency matrix that favors intra-document dominance of a word over intra-corpus dominance. However, it does not produce any knowledge about the term beyond the occurrence frequency. Topic Modelling (TM) is an alternative and widely investigated approach, which attempts to discover the underlying thematic structure within a text corpus as derived from co-occurrences of words across the documents \cite{MarkBelford}. A Topic Model typically consists of \(k\) topics, each represented by a ranked list of strongly-associated terms/words. Each topic represents trend or theme of the contents of the document. Belford et al. \cite{MarkBelford} extended TM in their \(E_{ns}TM\). They built on evidence by Topchy et al. \cite{Topchy1524981} that ensemble procedures encourage diversity and improve quality by integrating results across multiple iterations of individual algorithms.\vspace{-1em}

\begin{table}[h]
\centering
\small
\noindent\begin{tabular}{ |c|c|c| } 
 \hline
Feature-Type & Features  \\ 
  \hline
 context & holiday-food, beginner-cook, week-night,   \\
  & inexpensive , 6-people-or-more, potluck \\ 
    \hline
 cuisine & italian, hawaiian, tex-mex, chinese, cajun \\ 
   \hline
    equipment & saucepan, thermomix, wok, dutch-oven  \\ 
   \hline
    cooking & few-steps-recipe, less-than-one-hour, fried, \\ 
     process & slow-cooked, marinated, 4-hours-or-more \\ 
   \hline
 ingredient & poultry, feta, spaghetti, ham, shredded-meat\\
 \hline
  category & risotto, lasagna, stew, appetizer, pot-roast\\
 \hline
   nutrition & high-calcium, low-cholesterol, egg-free\\
 \hline
\end{tabular}
\caption{Summary\textsuperscript{\ref{ftn:note1}} of the extracted features from ETM }
\label{table:Table 1}
\end{table}\vspace{-2em}

To extract a set of significant features from our recipe corpus, we proceeded with \(E_{ns}TM\) \cite{MarkBelford} based on the generation and integration of the results produced by 100 runs of TM based on non-negative matrix factorization \cite{Koren:2009:MFT:1608565.1608614}. This produced a Topic-Term Weight Matrix where each column is a topic and each row determines the level of association between \{{Topic, Term}\} pair. To achieve a diverse and novel feature set we selected the top 30 topics and top 15 terms within each of these topics. Some terms appeared over multiple topics as they are involved in multiple food-trends. We consider the value of each \{{Topic, Term}\} pair in the Topic-Term Weight Matrix as the significance weight \(w_i\) for each term \(i\) within the corresponding topic. For terms existing over multiple topics we assigned \(w_i\) as the cumulative sum of their weight over all the corresponding topics. This produced a final set of 288 unique terms representing diverse aspects of food, e.g. cooking-approach, ingredient, equipment, serving and preservation techniques, context. These 288 terms, summarized in Table 1, are our identified Food Features and their corresponding weight are the proposed Feature Scores\footnote{\label{ftn:note1}The complete set of 288 features, their corresponding weights and set of food features correlated to healthier lifestyle are  available at https://github.com/MAK273/SupportingFileForHealthRecsys2019}.\vspace{0mm}

In this work, we adopted a simple recipe-to-feature relationship by representing each recipe as a vector of 288 features, where each feature value corresponds to its TF-IDF within the recipe. The transformation of the recipe corpus into a recipe-to-feature matrix, as shown in figure 1, reduces the bulk overload of food data while still holding enough information to retrieve each recipe.

\noindent\begin{minipage}{1.25in}
\label{fig:transform}
\small
\noindent\begin{tabular}{ |c|c|c| } 
 \hline
 \(Recipes\) & \(Plaintext\)  \\ 
  \hline
 \(R_1\) & \(Document_1\) \\ 
    \hline
 \(R_2\) & \(Document_2\) \\ 
   \hline
    \(.\) & \(......\) \\ 
   \hline
 \(R_n\) & \(Document_n\) \\ 
 \hline
\end{tabular}
\end{minipage}
\noindent\begin{minipage}{.5in}
\noindent$\xrightarrow{EnsTM}$
\end{minipage}
\noindent\begin{minipage}{1.25in}
\small
\noindent\begin{tabular}{ |c|c|c|c|c|c| } 
 \hline
 \(\) & \(f_1\) & \(f_2\) & . & . & \(f_{288}\) \\ 
  \hline
 \(R_1\) & \(0.79\)& \(0\)& . & . & \(.31\)\\ 
    \hline
 \(R_2\) & \(0\)& \(0\)& . & . & \(0\)\\ 
   \hline
 \(.\) & \(.\)& \(.\)& . & . & \(.\)\\ 
   \hline
 \(R_n\) & \(0.61\)& \(1\)& . & . & \(.08\)\\ 
 \hline
\end{tabular}
\end{minipage}\vspace{-2em}
\begin{figure}[h]
\caption{Recipe plain-text to feature vector transformation}
\end{figure}
\vspace{-0.8em}
\par\vspace{-.5em}

In the next step we used the identified food-features to learn user's preference. During their initial interaction with our FRS, users are asked to choose features with a like or dislike. (Note there was no requirement for users to rate all 288 features). To build the user-to-feature matrix the FRS assigns +5 to liked features, -5 to disliked features and 0 to any feature that has not been selected by the corresponding user. Unlike typical RS approaches we assigned an extreme negative value to disliked features. This was an important design decision and was done with the view to producing insights beyond user's food preferences, by enabling our system to implicitly capture important considerations such as nutritional restrictions or foods which users deliberately avoid.\vspace{0mm}

We implemented three \(E_{ns}TM\) based recommendation algorithms: FFbR, WFFbR, and FFbCF. The algorithms are named based on the attributes they operate on and the fundamental RS strategies they use. Each uses the recipe-to-feature matrix to transform user's positive and negative scores on features to user's scores on recipes.
\begin{itemize}
 \item Food Feature based Recommender (FFbR): This strategy assigns a preference score \(P\) for user \(u_a\) on a target recipe \(r_n\) based on the cumulative sum of \(u_a\)'s rating (dis/like) for all features \(f_{i(1,2,..,m)}\) present in \(r_n\). Where \(m\) is the total number features consisting \(r_n\). 
 \begin{equation} \label{eq:1}
 P(u_a,r_n) =
  \Bigg(\sum_{i=0}^{m}f_{i,u_a}\Bigg)^{'(0,5)}
\end{equation}
 
Instead of taking an average, we normalized the cumulative sum to a range \{{0 to 5}\} to favor recipes with more liked features than others. FFbR treats all food-features equally, assuming that each feature has an equal impact on user preferences.
 
 \item Weighted Food Feature based Recommender (WFFbR): With WFFbR we aimed to account for the differing impact of different food features. It scales \(u_a\)'s preference on a feature \(f_b\) with its corresponding feature score \(w_b\) and predicts \(u_a\)'s preference on \(r_n\) as the cumulative sum of the weighted preferences on all \(m\) features within \(r_n\).
  \begin{equation} \label{eq:2}
 P(u_a,r_n) =
  \Bigg(\sum_{i=0}^{m}f_{i,u_a} \times w_i \Bigg)^{'(0,5)}
\end{equation}
\item Food Feature based Collaborative Filtering (FFbCF): FFbCF applies the CF proposed by Freyne et al. \cite{Freyne:2010:IFP:1719970.1720021} in order to increase the knowledge on user's preference and predict user's preference score on food-features not been liked or disliked by the user. When user \(u_a\) first interacts with it the FFbCF identifies \(u_a\)'s  nearest neighbors based on similar ratings on overlapping features. We implemented KNN clustering \cite{KNNCover:2006:NNP:2263261.2267456} to identify top \(n\) nearest neighbours of \(u_a\). For a new feature \(f_b\) FFbCF predicted \(u_a\)'s preference as,
\vspace{-1em}

  \begin{equation} \label{eq:3}
 P(f_{b,u_a}) = 
  \dfrac{\sum_{i=0}^{n}f_{b,i}}{n}
\end{equation}

With this more densely populated user-to-feature matrix FFbCF generates \( P(u_a,r_n) \) for using equation \ref{eq:1}.
\end{itemize}
\section{EVALUATION}
In order to test the \(E_{ns}TM\) base FRS strategies, we conducted a user study with 48 users of varying nationality, ethnicity, gender and age. Participants were recruited though social media groups within UCD. All participants were entered into a draw for a 50\EUR{} gift voucher. Ethics permission for this study was provided by UCD office of research ethics.\vspace{0mm}

A smaller recipe-corpus of 92,539 recipes with valid images was used as the primary recipe data-set. The study compared four approaches: the three \(E_{ns}TM\) based FRS strategies and a CB approach. Each approach predicted user's preference on all 92,539 recipes. For each recommendation strategy, the top 2,100 recipes with highest prediction score were divided into 7 equal sized epochs and from each epoch one recipe was randomly selected. This approach was taken to support diversity and allow users to have more options at their disposal.\vspace{0mm}

We developed a website\footnote{\label{ftn:note2}Demo of the website could be found at https://youtu.be/ujaB0FiqRwk}.\vspace{0mm} and hosted it under the university domain. Participants were first required to access the website and indicate their informed consent and then create a user-name and password. They could then log into a secure website that displayed an interactive panel of images representing all 288 features, in the order of their feature weight. They were asked to select at least 20 features which they like and at least 20 features which they dislike. This information was used to create a user profile. Once created, participants could log into their profile and browse the features to update their likes and dislikes. Participants also selected an appointment time for the main experiment.

During the main experiment participants were shown a series of four recommendation lists corresponding to each of our recommendation algorithms. Each list consisted of seven recipes. The order in which the recommendation lists were presented was fully counter-balanced across the 48 participants. Within each list, participants were required to rate each individual recipe on a 5 star rating scale, where 0 and 5 represented "not like at all" and "liked very much" respectively.   

\subsection {RESULTS}
\textbf{Accuracy}: The accuracy of the recommendations has been evaluated based on participant ratings of recipes. For each participant, the average rating across the seven-item list generated by each recommendation strategy was calculated. Figure \ref{fig:medianbox} shows the mean score of each algorithm across all users. Again the pure CB approach was the poorest performer. This was confirmed though statistical analysis. We first conducted a repeated measures analysis of variance that compared the mean ratings of participants across the four algorithms. The result, F(3,188)= 14.42229, p<0.001, indicates a significant difference within the results. Paired sample t-tests were then conducted between the individual algorithms, with a null hypothesis in each case of no difference in the mean ratings. We do not find a significant difference between participants ratings across the \(E_{ns}TM\) approaches, indicating that they all performed equally well in terms of accuracy. There was however a significant difference in participants ratings between each of the \(E_{ns}TM\) approaches and the CB baseline, with p < 0.001 in each case. This suggests that each \(E_{ns}TM\) based approach performed significantly better than the baseline CB approach.\vspace{-.5em} 
\noindent\begin{figure}[h!]
\centering
\tiny
\begin{tikzpicture}
\filldraw[ball color=red!80,shading=ball] (3.8,.65) circle
        (0.06cm) node[right]{3.42};
        \filldraw[ball color=red!80,shading=ball] (3.65,1.5) circle
        (0.06cm) node[right]{3.33};
          \filldraw[ball color=red!80,shading=ball] (3.8,2.5) circle
        (0.06cm) node[right]{3.45};
        \filldraw[ball color=red!80,shading=ball] (2.8,3.4) circle
        (0.06cm) node[right]{2.8};
\begin{axis}[boxplot/draw direction=x,                 
             yticklabels={Test A, Test B ,FFbCF, WFFbR , FFbR, CB},
            x tick label style={enlargelimits=0.05},
            xtick={1,1.5,2,2.5,3,3.5,4,4.5,5},
            xticklabels={1,1.5,2,2.5,3,3.5,4,4.5,5},
            height=5.5cm,
            width=8cm,
            ]
\pgfplotstablegetrowsof{\datatable}
\pgfmathtruncatemacro\TotalRows{\pgfplotsretval-1}
\pgfplotsinvokeforeach{0,...,\TotalRows}
{  
  \addplot+[
  boxplot prepared from table={
    table=\datatable,
    row=#1,
    lower whisker=lw,
    upper whisker=uw,
    lower quartile=lq,
    upper quartile=uq,
    median=med
  },
  boxplot prepared,
  area legend
  ]
  coordinates {};
 
}
\end{axis}
\end{tikzpicture}
\caption{Cumulative preference score from each user}
\label{fig:medianbox}
\end{figure}
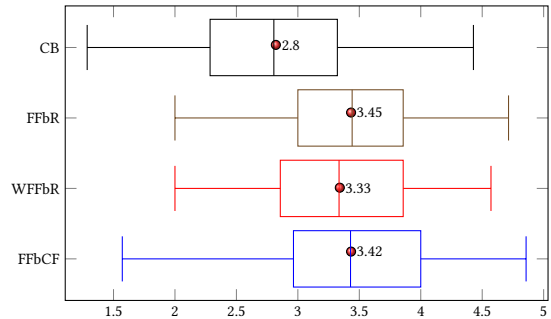\vspace{-0.5em}


\textbf{Coverage}: Here we consider the coverage achieved by each algorithm across all users, that is, the percentage of recipe-user pairs where the algorithm was able to generate a prediction. The notable outlier is CB, which produced coverage of only 20\%. FFbR and WFFbR both had user's preferences for an average of 51 of our 288 features and both produced a coverage of 91.57\%. FFbCF, with a more densely populated user-to-feature matrix, provided 100\% coverage, with predictions for all recipe-user pairs.

\begin{figure}[h]
\centering
\includegraphics[ height=4cm,width=8cm]{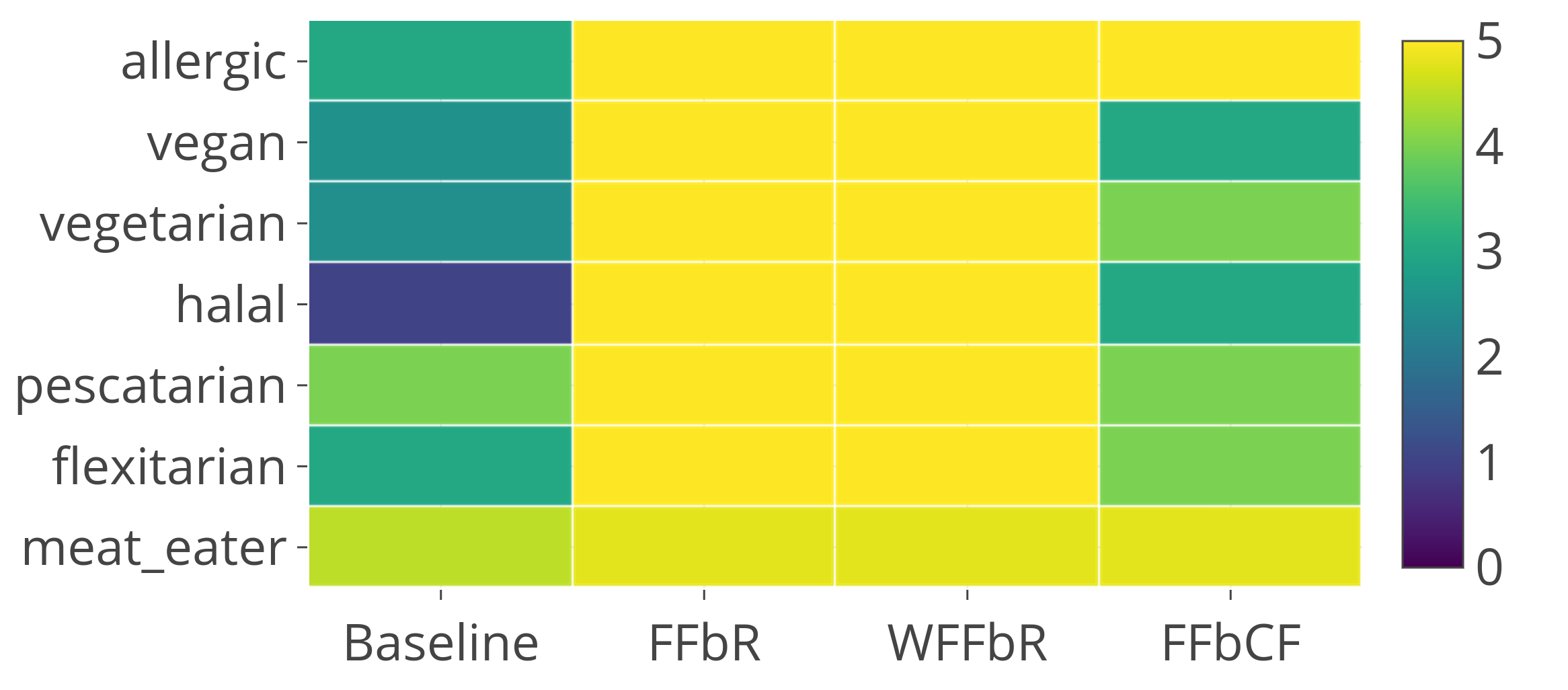}
    \caption{How successfully each method identified user's food practice}
    \label{fig:foodpractices}
\end{figure}

\textbf{Implicitly capturing food practices}: Another practical aspect of knowledge building for a FRS is an algorithm's ability to predict important aspects of a user's food practices from available user information. For example, while both vegetarians and vegans eat vegetables, eggs should only be recommended to vegetarians. Figure \ref{fig:foodpractices} shows that the CB baseline performed poorly in this regard. In contrast FFbR , WFFbR identified user's food practice 100\% accurately. Here the feature-to-recipe direct relationship extends the dislike property of the FRS as an effective identifier tool. The reason FFbCF failed to predict food practice for some users is the collaborative effect of their neighbour's food practice.
\vspace{-1em}
\noindent\vspace{-1.0em}

\textbf{Correlation between lifestyle and food-features}: Further analysis on the data-set collected from the user study exposed interesting associations between users' lifestyle and their feature-preference. Users were categorized under different health-groups based on three different healthiness measures: activity\char`_level, BMI and average food\char`_healthScore. User's activity\char`_level was a self reported assessment by user. BMI was calculated from users' height and weight \cite{website:WHO-BMI}. User's average food\char`_healthScore was defined as the average health-score \cite{Harvey_learninguser} of all recipes user liked (rated 4 or more). Table \ref{table:classLavel} summarizes the category labels corresponding to each healthiness measure and the guideline associated with each categorization criteria. The activity\char`_level and food\char`_healthScore based categorization showed agreement on the healthiness of user's lifestyle preference. Users from more active groups (moderately\char`_active, extra\char`_active) were also identified in higher food\char`_healthScore groups (moderately\char`_healthy, very\char`_healthy). Where, BMI based categorization was not predictive of either of activity\char`_level and food\char`_healthScore based categorization.

\vspace{-1.0em}

\begin{table}[h!]
\centering
\small
\noindent\begin{tabular}{ |c|c|c| } 
 \hline
  Scale & Guideline & User Group   \\ 
  \hline
  Activity &FAO: activity level,  & sedentary, lightly\char`_active,\\
  level & energy intake   \cite{website:FAO_activitylevelClasses} & moderately\char`_active, extra\char`_active \\ 
    \hline
  BMI &  WHO:BMI\cite{website:WHO-BMI} & underweight, normal\char`_weight ,\\
  & &  pre\char`_obesity, obesityclass\char`_1	 \\ 
   \hline
  Food &FSA: nutrient  & less\char`_healthy, moderately\char`_healthy, \\
 choices  & intake guideline \cite{website:FSA:Nutrient} & very\char`_healthy  \\ 
  \hline
\end{tabular}
\caption{user-groups based on different health variable.  }
\label{table:classLavel}
\end{table}

\vspace{-2em}
The aim of the categorization was to investigate, if there is any pattern in the interactions between certain health-group and any food features.
Finding the correlation between these two variables allows us to assess whether healthier users tend to like or dislike a particular feature. A natural approach for such analysis is the application of machine learning classification algorithms to access the predictive capabilities of these features, although due the small sample size (48 users) and the high degree of imbalance in the class size across all three scales, a simple correlation analysis is used in favour of these methods in this instance.

Results expressed interesting associations between health-groups and features. Given that the group/category-level associated with activity\char`_level and food\char`_healthScore are ordinal in nature, we conducted a Spearman rank correlation analysis to find the degree of association between preference (positive/negative) for features and health-groups. Table \ref{table:corelation1} shows the strongest significant features with p<0.05 for a sample of 48 users.
Features popular among healthier user group have the potential to be leveraged as initial recommendation for new users who are looking for inspiration on healthier food-ideas or recipes.\vspace{-2em}

\begin{table}[h]
\begin{tabular}{llll}
\multicolumn{4}{c}{}                                                                                                                              \\ \hline
\multicolumn{2}{|c|}{Average Food HealthScore}                        & \multicolumn{2}{c|}{Activity Level}                                       \\ \hline
\multicolumn{1}{|l|}{Feature}         & \multicolumn{1}{l|}{r}        & \multicolumn{1}{l|}{Feature}             & \multicolumn{1}{l|}{r}         \\ \hline
\multicolumn{1}{|l|}{peanut-butter}   & \multicolumn{1}{l|}{0.447989} & \multicolumn{1}{l|}{wing}                & \multicolumn{1}{l|}{0.441152}  \\ \hline
\multicolumn{1}{|l|}{granola}         & \multicolumn{1}{l|}{0.365171} & \multicolumn{1}{l|}{tuna}                & \multicolumn{1}{l|}{0.430467}  \\ \hline
\multicolumn{1}{|l|}{lentil}          & \multicolumn{1}{l|}{0.360767} & \multicolumn{1}{l|}{tilapia}             & \multicolumn{1}{l|}{0.363502}  \\ \hline
\multicolumn{1}{|l|}{indian}          & \multicolumn{1}{l|}{0.356347} & \multicolumn{1}{l|}{salmon}              & \multicolumn{1}{l|}{0.359852}  \\ \hline
\multicolumn{1}{|l|}{cauliflower}     & \multicolumn{1}{l|}{0.352353} & \multicolumn{1}{l|}{hawaiian}            & \multicolumn{1}{l|}{0.346401}  \\ \hline
\multicolumn{1}{|l|}{low-cholesterol} & \multicolumn{1}{l|}{0.350818} & \multicolumn{1}{l|}{canadian}            & \multicolumn{1}{l|}{0.322470}  \\ \hline
\multicolumn{1}{|l|}{maple}           & \multicolumn{1}{l|}{0.321131} & \multicolumn{1}{l|}{smoothy}             & \multicolumn{1}{l|}{0.314174}  \\ \hline
\multicolumn{1}{|l|}{vegetable}       & \multicolumn{1}{l|}{0.307459}  & \multicolumn{1}{l|}{chicken-thighs-legs} & \multicolumn{1}{l|}{0.314059} \\ \hline
\multicolumn{1}{|l|}{wheat}           & \multicolumn{1}{l|}{0.303326} & \multicolumn{1}{l|}{halibut}             & \multicolumn{1}{l|}{0.310990}  \\ \hline
\multicolumn{1}{|l|}{carrot}          & \multicolumn{1}{l|}{0.303052} & \multicolumn{1}{l|}{main-dish}           & \multicolumn{1}{l|}{0.303345}  \\ \hline
\end{tabular}
\caption{ Top\textsuperscript{\ref{ftn:note1}} 10 positively correlated features to healthier user-groups}
\label{table:corelation1}
\end{table}\vspace{-3.0em}
\section{CONCLUSIONS AND FUTURE WORK}
This work presents an initial evaluation of \(E_{ns}TM\) based FRS. Results show that \(E_{ns}TM\) based approaches performs significantly better than a conventional CB approach. It provides a universal feature extraction approach that can generate a set of significant food-features from any recipe/ menu/ food corpus. The features have the added advantage of being human understandable and allowed us to directly model user preferences. \(E_{ns}TM\) based feature identification resolves the limitation of user-group dependency and is capable of making food recommendations for users from diverse nationality, ethnicity and culture. It allows for the generation of recommendations without the need for existing user ratings on recipes, helping to address the cold start problem. By working with a reduced feature set, \(E_{ns}TM\) also enables computationally efficient recommendation. Furthermore the the subset of nutritional features within our food features supports the proposed \(E_{ns}TM\) approaches to  personalize the Reclist according user's nutritional preference. 

While there was no significant difference between the three \(E_{ns}TM\) based approaches in terms of users' recipe ratings, the use of \(E_{ns}TM\) in combination with CF provided best coverage, predicting user preferences across 100\% of our recipe corpus. However, the CF based approach performed more poorly in terms of implicit understanding of users' food practices. In future work we aim to
focus on applying the \(E_{ns}TM\) based recommenders to support diet/menu planning by incorporating health-aware filtering strategies, with the view to providing long-term, guided and healthier food choices. The positive and negative popularity of features among certain health-groups also inspired us to investigate food feature in comparison with healthiness clues for user modeling and recipe recommendation.\vspace{-1.0em}

%
\bibliographystyle{ACM-Reference-Format}
\bibliography{article}


\begin{thebibliography}{33}


\ifx \showCODEN    \undefined \def \showCODEN     #1{\unskip}     \fi
\ifx \showDOI      \undefined \def \showDOI       #1{#1}\fi
\ifx \showISBNx    \undefined \def \showISBNx     #1{\unskip}     \fi
\ifx \showISBNxiii \undefined \def \showISBNxiii  #1{\unskip}     \fi
\ifx \showISSN     \undefined \def \showISSN      #1{\unskip}     \fi
\ifx \showLCCN     \undefined \def \showLCCN      #1{\unskip}     \fi
\ifx \shownote     \undefined \def \shownote      #1{#1}          \fi
\ifx \showarticletitle \undefined \def \showarticletitle #1{#1}   \fi
\ifx \showURL      \undefined \def \showURL       {\relax}        \fi
\providecommand\bibfield[2]{#2}
\providecommand\bibinfo[2]{#2}
\providecommand\natexlab[1]{#1}
\providecommand\showeprint[2][]{arXiv:#2}

\bibitem[\protect\citeauthoryear{??}{web}{[n. d.]a}]%
        {website:FSA:Nutrient}
 \bibinfo{year}{[n. d.]}\natexlab{a}.
\newblock \bibinfo{title}{FSA Nutrient and Food Guidelines}.
\newblock
\newblock
\urldef\tempurl%
\url{https://www.ptdirect.com/training-design/nutrition/national-nutrition-guidelines-united-kingdom}
\showURL{%
\tempurl}
\newblock
\shownote{Accessed : March 2018.}


\bibitem[\protect\citeauthoryear{??}{web}{[n. d.]b}]%
        {website:geniuskitchen}
 \bibinfo{year}{[n. d.]}\natexlab{b}.
\newblock \bibinfo{title}{Geniuskitchen}.
\newblock
\newblock
\urldef\tempurl%
\url{http://www.geniuskitchen.com}
\showURL{%
\tempurl}
\newblock
\shownote{Accessed : March 2018.}


\bibitem[\protect\citeauthoryear{??}{web}{2009a}]%
        {website:FAO_activitylevelClasses}
 \bibinfo{year}{2009}\natexlab{a}.
\newblock \bibinfo{title}{FAO energy requirement guideline}.
\newblock
\newblock
\urldef\tempurl%
\url{http://www.fao.org/3/y5686e/y5686e07.htm}
\showURL{%
\tempurl}
\newblock
\shownote{Accessed :March 2018.}


\bibitem[\protect\citeauthoryear{??}{web}{2009b}]%
        {website:WHO-BMI}
 \bibinfo{year}{2009}\natexlab{b}.
\newblock \bibinfo{title}{WHO : Body mass index}.
\newblock
\newblock
\urldef\tempurl%
\url{http://www.euro.who.int/en/health-topics/disease-prevention/nutrition/a-healthy-lifestyle/body-mass-index-bmi}
\showURL{%
\tempurl}
\newblock
\shownote{Accessed :March 2018.}


\bibitem[\protect\citeauthoryear{??}{web}{2018}]%
        {website:WHO-fact-sheet}
 \bibinfo{year}{2018}\natexlab{}.
\newblock \bibinfo{title}{WHO fact sheet. 2018.}
\newblock
\newblock
\urldef\tempurl%
\url{http://www.who.int/mediacentre/factsheets/fs311/en/}
\showURL{%
\tempurl}
\newblock
\shownote{Accessed :March 2018.}


\bibitem[\protect\citeauthoryear{A~Bisogni, Jastran, Seligson, and
  Thompson}{A~Bisogni et~al\mbox{.}}{2012}]%
        {article:HowPeopleInterpretHealthyEating}
\bibfield{author}{\bibinfo{person}{Carole A~Bisogni}, \bibinfo{person}{Margaret
  Jastran}, \bibinfo{person}{Marc Seligson}, {and} \bibinfo{person}{Alyssa
  Thompson}.} \bibinfo{year}{2012}\natexlab{}.
\newblock \showarticletitle{How People Interpret Healthy Eating: Contributions
  of Qualitative Research}.
\newblock \bibinfo{journal}{\emph{Journal of nutrition education and behavior}}
   \bibinfo{volume}{44} (\bibinfo{date}{07} \bibinfo{year}{2012}),
  \bibinfo{pages}{282--301}.
\newblock
\urldef\tempurl%
\url{https://doi.org/10.1016/j.jneb.2011.11.009}
\showDOI{\tempurl}


\bibitem[\protect\citeauthoryear{Belford, MacNamee, and Greene}{Belford
  et~al\mbox{.}}{2016}]%
        {MarkBelford}
\bibfield{author}{\bibinfo{person}{Mark Belford}, \bibinfo{person}{Brian
  MacNamee}, {and} \bibinfo{person}{Derek Greene}.}
  \bibinfo{year}{2016}\natexlab{}.
\newblock \bibinfo{title}{Ensemble Topic Modeling via Matrix Factorization}.
\newblock
\newblock


\bibitem[\protect\citeauthoryear{Bobadilla, Ortega, Hernando, and
  Bernal}{Bobadilla et~al\mbox{.}}{2012}]%
        {Bobadilla:2012:CFA:2107778.2107834}
\bibfield{author}{\bibinfo{person}{Jes\'{u}S Bobadilla},
  \bibinfo{person}{Fernando Ortega}, \bibinfo{person}{Antonio Hernando}, {and}
  \bibinfo{person}{Jes\'{u}S Bernal}.} \bibinfo{year}{2012}\natexlab{}.
\newblock \showarticletitle{A Collaborative Filtering Approach to Mitigate the
  New User Cold Start Problem}.
\newblock \bibinfo{journal}{\emph{Know.-Based Syst.}}  \bibinfo{volume}{26}
  (\bibinfo{date}{Feb.} \bibinfo{year}{2012}), \bibinfo{pages}{225--238}.
\newblock
\showISSN{0950-7051}
\urldef\tempurl%
\url{https://doi.org/10.1016/j.knosys.2011.07.021}
\showDOI{\tempurl}


\bibitem[\protect\citeauthoryear{Cover and Hart}{Cover and Hart}{2006}]%
        {KNNCover:2006:NNP:2263261.2267456}
\bibfield{author}{\bibinfo{person}{T. Cover} {and} \bibinfo{person}{P. Hart}.}
  \bibinfo{year}{2006}\natexlab{}.
\newblock \showarticletitle{Nearest Neighbor Pattern Classification}.
\newblock \bibinfo{journal}{\emph{IEEE Trans. Inf. Theor.}}
  \bibinfo{volume}{13}, \bibinfo{number}{1} (\bibinfo{date}{Sept.}
  \bibinfo{year}{2006}), \bibinfo{pages}{21--27}.
\newblock
\showISSN{0018-9448}
\urldef\tempurl%
\url{https://doi.org/10.1109/TIT.1967.1053964}
\showDOI{\tempurl}


\bibitem[\protect\citeauthoryear{Elsweiler and Harvey}{Elsweiler and
  Harvey}{2015}]%
        {Elsweiler:2015:TAM:2792838.2799665}
\bibfield{author}{\bibinfo{person}{David Elsweiler} {and}
  \bibinfo{person}{Morgan Harvey}.} \bibinfo{year}{2015}\natexlab{}.
\newblock \showarticletitle{Towards Automatic Meal Plan Recommendations for
  Balanced Nutrition}. In \bibinfo{booktitle}{\emph{Proceedings of the 9th ACM
  Conference on Recommender Systems}} \emph{(\bibinfo{series}{RecSys '15})}.
  \bibinfo{publisher}{ACM}, \bibinfo{address}{New York, NY, USA},
  \bibinfo{pages}{313--316}.
\newblock
\showISBNx{978-1-4503-3692-5}
\urldef\tempurl%
\url{https://doi.org/10.1145/2792838.2799665}
\showDOI{\tempurl}


\bibitem[\protect\citeauthoryear{Elsweiler, Trattner, and Harvey}{Elsweiler
  et~al\mbox{.}}{2017}]%
        {Elsweiler:EFC:3077136:3080826}
\bibfield{author}{\bibinfo{person}{David Elsweiler}, \bibinfo{person}{Christoph
  Trattner}, {and} \bibinfo{person}{Morgan Harvey}.}
  \bibinfo{year}{2017}\natexlab{}.
\newblock \showarticletitle{Exploiting Food Choice Biases for Healthier Recipe
  Recommendation}. In \bibinfo{booktitle}{\emph{Proceedings of the 40th
  International ACM SIGIR Conference on Research and Development in Information
  Retrieval}} \emph{(\bibinfo{series}{SIGIR '17})}. \bibinfo{publisher}{ACM},
  \bibinfo{address}{New York, NY, USA}, \bibinfo{pages}{575--584}.
\newblock
\showISBNx{978-1-4503-5022-8}
\urldef\tempurl%
\url{https://doi.org/10.1145/3077136.3080826}
\showDOI{\tempurl}


\bibitem[\protect\citeauthoryear{Freyne and Berkovsky}{Freyne and
  Berkovsky}{2010a}]%
        {Freyne:2010:IFP:1719970.1720021}
\bibfield{author}{\bibinfo{person}{Jill Freyne} {and} \bibinfo{person}{Shlomo
  Berkovsky}.} \bibinfo{year}{2010}\natexlab{a}.
\newblock \showarticletitle{Intelligent Food Planning: Personalized Recipe
  Recommendation}. In \bibinfo{booktitle}{\emph{Proceedings of the 15th
  International Conference on Intelligent User Interfaces}}
  \emph{(\bibinfo{series}{IUI '10})}. \bibinfo{publisher}{ACM},
  \bibinfo{address}{New York, NY, USA}, \bibinfo{pages}{321--324}.
\newblock
\showISBNx{978-1-60558-515-4}
\urldef\tempurl%
\url{https://doi.org/10.1145/1719970.1720021}
\showDOI{\tempurl}


\bibitem[\protect\citeauthoryear{Freyne and Berkovsky}{Freyne and
  Berkovsky}{2010b}]%
        {Freyne:2010:RFR:2149528.2149568}
\bibfield{author}{\bibinfo{person}{Jill Freyne} {and} \bibinfo{person}{Shlomo
  Berkovsky}.} \bibinfo{year}{2010}\natexlab{b}.
\newblock \showarticletitle{Recommending Food: Reasoning on Recipes and
  Ingredients}. In \bibinfo{booktitle}{\emph{Proceedings of the 18th
  International Conference on User Modeling, Adaptation, and Personalization}}
  \emph{(\bibinfo{series}{UMAP'10})}. \bibinfo{publisher}{Springer-Verlag},
  \bibinfo{address}{Berlin, Heidelberg}, \bibinfo{pages}{381--386}.
\newblock
\showISBNx{3-642-13469-6, 978-3-642-13469-2}
\urldef\tempurl%
\url{https://doi.org/10.1007/978-3-642-13470-8_36}
\showDOI{\tempurl}


\bibitem[\protect\citeauthoryear{Ge, Elahi, Ferna\'{a}ndez-Tob\'{\i}as, Ricci,
  and Massimo}{Ge et~al\mbox{.}}{2015a}]%
        {Ge:2015:UTL:2750511.2750528}
\bibfield{author}{\bibinfo{person}{Mouzhi Ge}, \bibinfo{person}{Mehdi Elahi},
  \bibinfo{person}{Ignacio Ferna\'{a}ndez-Tob\'{\i}as},
  \bibinfo{person}{Francesco Ricci}, {and} \bibinfo{person}{David Massimo}.}
  \bibinfo{year}{2015}\natexlab{a}.
\newblock \showarticletitle{Using Tags and Latent Factors in a Food Recommender
  System}. In \bibinfo{booktitle}{\emph{Proceedings of the 5th International
  Conference on Digital Health 2015}} \emph{(\bibinfo{series}{DH '15})}.
  \bibinfo{publisher}{ACM}, \bibinfo{address}{New York, NY, USA},
  \bibinfo{pages}{105--112}.
\newblock
\showISBNx{978-1-4503-3492-1}
\urldef\tempurl%
\url{https://doi.org/10.1145/2750511.2750528}
\showDOI{\tempurl}


\bibitem[\protect\citeauthoryear{Ge, Ricci, and Massimo}{Ge
  et~al\mbox{.}}{2015b}]%
        {Ge:2015:HFR:2792838.2796554}
\bibfield{author}{\bibinfo{person}{Mouzhi Ge}, \bibinfo{person}{Francesco
  Ricci}, {and} \bibinfo{person}{David Massimo}.}
  \bibinfo{year}{2015}\natexlab{b}.
\newblock \showarticletitle{Health-aware Food Recommender System}. In
  \bibinfo{booktitle}{\emph{Proceedings of the 9th ACM Conference on
  Recommender Systems}} \emph{(\bibinfo{series}{RecSys '15})}.
  \bibinfo{publisher}{ACM}, \bibinfo{address}{New York, NY, USA},
  \bibinfo{pages}{333--334}.
\newblock
\showISBNx{978-1-4503-3692-5}
\urldef\tempurl%
\url{https://doi.org/10.1145/2792838.2796554}
\showDOI{\tempurl}


\bibitem[\protect\citeauthoryear{Gu and Wang}{Gu and Wang}{2009}]%
        {Gu:2009:CFB:1701835.1701851}
\bibfield{author}{\bibinfo{person}{Hanshen Gu} {and} \bibinfo{person}{Dong
  Wang}.} \bibinfo{year}{2009}\natexlab{}.
\newblock \showarticletitle{A Content-aware Fridge Based on RFID in Smart Home
  for Home-healthcare}. In \bibinfo{booktitle}{\emph{Proceedings of the 11th
  International Conference on Advanced Communication Technology - Volume 2}}
  \emph{(\bibinfo{series}{ICACT'09})}. \bibinfo{publisher}{IEEE Press},
  \bibinfo{address}{Piscataway, NJ, USA}, \bibinfo{pages}{987--990}.
\newblock
\showISBNx{978-8-9551-9138-7}
\urldef\tempurl%
\url{https://doi.org/citation.cfm?id=1701835.1701851}
\showDOI{\tempurl}


\bibitem[\protect\citeauthoryear{Harvey, Ludwig, and Elsweiler}{Harvey
  et~al\mbox{.}}{[n. d.]}]%
        {Harvey_learninguser}
\bibfield{author}{\bibinfo{person}{Morgan Harvey}, \bibinfo{person}{Bernd
  Ludwig}, {and} \bibinfo{person}{David Elsweiler}.} \bibinfo{year}{[n.
  d.]}\natexlab{}.
\newblock \bibinfo{title}{Learning user tastes: a first step to generating
  healthy meal plans?}
\newblock
\newblock


\bibitem[\protect\citeauthoryear{Harvey, Ludwig, and Elsweiler}{Harvey
  et~al\mbox{.}}{2013}]%
        {Harvey:2013:YYE:2651320.2651339}
\bibfield{author}{\bibinfo{person}{Morgan Harvey}, \bibinfo{person}{Bernd
  Ludwig}, {and} \bibinfo{person}{David Elsweiler}.}
  \bibinfo{year}{2013}\natexlab{}.
\newblock \showarticletitle{You Are What You Eat: Learning User Tastes for
  Rating Prediction}. In \bibinfo{booktitle}{\emph{Proceedings of the 20th
  International Symposium on String Processing and Information Retrieval -
  Volume 8214}} \emph{(\bibinfo{series}{SPIRE 2013})}.
  \bibinfo{publisher}{Springer-Verlag}, \bibinfo{address}{Berlin, Heidelberg},
  \bibinfo{pages}{153--164}.
\newblock
\showISBNx{978-3-319-02431-8}
\urldef\tempurl%
\url{https://doi.org/10.1007/978-3-319-02432-5_19}
\showDOI{\tempurl}


\bibitem[\protect\citeauthoryear{Koren, Bell, and Volinsky}{Koren
  et~al\mbox{.}}{2009}]%
        {Koren:2009:MFT:1608565.1608614}
\bibfield{author}{\bibinfo{person}{Yehuda Koren}, \bibinfo{person}{Robert
  Bell}, {and} \bibinfo{person}{Chris Volinsky}.}
  \bibinfo{year}{2009}\natexlab{}.
\newblock \showarticletitle{Matrix Factorization Techniques for Recommender
  Systems}.
\newblock \bibinfo{journal}{\emph{Computer}} \bibinfo{volume}{42},
  \bibinfo{number}{8} (\bibinfo{date}{Aug.} \bibinfo{year}{2009}),
  \bibinfo{pages}{30--37}.
\newblock
\showISSN{0018-9162}
\urldef\tempurl%
\url{https://doi.org/10.1109/MC.2009.263}
\showDOI{\tempurl}


\bibitem[\protect\citeauthoryear{Kuo, Li, Shan, and Lee}{Kuo
  et~al\mbox{.}}{2012}]%
        {Kuo:2012:IMP:2390776.2390778}
\bibfield{author}{\bibinfo{person}{Fang-Fei Kuo}, \bibinfo{person}{Cheng-Te
  Li}, \bibinfo{person}{Man-Kwan Shan}, {and} \bibinfo{person}{Suh-Yin Lee}.}
  \bibinfo{year}{2012}\natexlab{}.
\newblock \showarticletitle{Intelligent Menu Planning: Recommending Set of
  Recipes by Ingredients}. In \bibinfo{booktitle}{\emph{Proceedings of the ACM
  Multimedia 2012 Workshop on Multimedia for Cooking and Eating Activities}}
  \emph{(\bibinfo{series}{CEA '12})}. \bibinfo{publisher}{ACM},
  \bibinfo{address}{New York, NY, USA}, \bibinfo{pages}{1--6}.
\newblock
\showISBNx{978-1-4503-1592-0}
\urldef\tempurl%
\url{https://doi.org/10.1145/2390776.2390778}
\showDOI{\tempurl}


\bibitem[\protect\citeauthoryear{Ramos}{Ramos}{2003}]%
        {articleTF_IDF}
\bibfield{author}{\bibinfo{person}{Juan Ramos}.}
  \bibinfo{year}{2003}\natexlab{}.
\newblock \showarticletitle{Using TF-IDF to determine word relevance in
  document queries}.
\newblock  (\bibinfo{date}{01} \bibinfo{year}{2003}).
\newblock


\bibitem[\protect\citeauthoryear{Rokicki, Herder, Ku\'{s}mierczyk, and
  Trattner}{Rokicki et~al\mbox{.}}{2016}]%
        {Rokicki:2016:PPG:2930238.2930248}
\bibfield{author}{\bibinfo{person}{Markus Rokicki}, \bibinfo{person}{Eelco
  Herder}, \bibinfo{person}{Tomasz Ku\'{s}mierczyk}, {and}
  \bibinfo{person}{Christoph Trattner}.} \bibinfo{year}{2016}\natexlab{}.
\newblock \showarticletitle{Plate and Prejudice: Gender Differences in Online
  Cooking}. In \bibinfo{booktitle}{\emph{Proceedings of the 2016 Conference on
  User Modeling Adaptation and Personalization}} \emph{(\bibinfo{series}{UMAP
  '16})}. \bibinfo{publisher}{ACM}, \bibinfo{address}{New York, NY, USA},
  \bibinfo{pages}{207--215}.
\newblock
\showISBNx{978-1-4503-4368-8}
\urldef\tempurl%
\url{https://doi.org/10.1145/2930238.2930248}
\showDOI{\tempurl}


\bibitem[\protect\citeauthoryear{Rokicki, Trattner, and Herder}{Rokicki
  et~al\mbox{.}}{2018}]%
        {Rokicki2018TheIO}
\bibfield{author}{\bibinfo{person}{Markus Rokicki}, \bibinfo{person}{Christoph
  Trattner}, {and} \bibinfo{person}{Eelco Herder}.}
  \bibinfo{year}{2018}\natexlab{}.
\newblock \showarticletitle{The Impact of Recipe Features, Social Cues and
  Demographics on Estimating the Healthiness of Online Recipes}. In
  \bibinfo{booktitle}{\emph{ICWSM}}.
\newblock


\bibitem[\protect\citeauthoryear{Sobecki, Babiak, and Slanina}{Sobecki
  et~al\mbox{.}}{2006}]%
        {Sobecki:2006:AHR:2165946.2166064}
\bibfield{author}{\bibinfo{person}{Janusz Sobecki}, \bibinfo{person}{E.
  Babiak}, {and} \bibinfo{person}{M. Slanina}.}
  \bibinfo{year}{2006}\natexlab{}.
\newblock \showarticletitle{Application of Hybrid Recommendation in Web-based
  Cooking Assistant}. In \bibinfo{booktitle}{\emph{Proceedings of the 10th
  International Conference on Knowledge-Based Intelligent Information and
  Engineering Systems - Volume Part III}} \emph{(\bibinfo{series}{KES'06})}.
  \bibinfo{publisher}{Springer-Verlag}, \bibinfo{address}{Berlin, Heidelberg},
  \bibinfo{pages}{797--804}.
\newblock
\showISBNx{3-540-46542-1, 978-3-540-46542-3}
\urldef\tempurl%
\url{https://doi.org/10.1007/11893011_101}
\showDOI{\tempurl}


\bibitem[\protect\citeauthoryear{Teng, Lin, and Adamic}{Teng
  et~al\mbox{.}}{2012}]%
        {Teng:2012:RRU:2380718:2380757}
\bibfield{author}{\bibinfo{person}{Chun-Yuen Teng}, \bibinfo{person}{Yu-Ru
  Lin}, {and} \bibinfo{person}{Lada~A. Adamic}.}
  \bibinfo{year}{2012}\natexlab{}.
\newblock \showarticletitle{Recipe Recommendation Using Ingredient Networks}.
  In \bibinfo{booktitle}{\emph{Proceedings of the 4th Annual ACM Web Science
  Conference}} \emph{(\bibinfo{series}{WebSci '12})}. \bibinfo{publisher}{ACM},
  \bibinfo{address}{New York, NY, USA}, \bibinfo{pages}{298--307}.
\newblock
\showISBNx{978-1-4503-1228-8}
\urldef\tempurl%
\url{https://doi.org/10.1145/2380718.2380757}
\showDOI{\tempurl}


\bibitem[\protect\citeauthoryear{Topchy, Jain, and Punch}{Topchy
  et~al\mbox{.}}{2005}]%
        {Topchy1524981}
\bibfield{author}{\bibinfo{person}{Alexander Topchy}, \bibinfo{person}{Anil~K.
  Jain}, {and} \bibinfo{person}{William~F. Punch}.}
  \bibinfo{year}{2005}\natexlab{}.
\newblock \showarticletitle{Clustering ensembles: models of consensus and weak
  partitions}.
\newblock \bibinfo{journal}{\emph{IEEE Transactions on Pattern Analysis and
  Machine Intelligence}} \bibinfo{volume}{27}, \bibinfo{number}{12}
  (\bibinfo{date}{Dec} \bibinfo{year}{2005}), \bibinfo{pages}{1866--1881}.
\newblock
\showISSN{0162-8828}
\urldef\tempurl%
\url{https://doi.org/10.1109/TPAMI.2005.237}
\showDOI{\tempurl}


\bibitem[\protect\citeauthoryear{Trattner and Elsweiler}{Trattner and
  Elsweiler}{2017a}]%
        {DBLP:journals/corr/abs-1711-02760}
\bibfield{author}{\bibinfo{person}{Christoph Trattner} {and}
  \bibinfo{person}{David Elsweiler}.} \bibinfo{year}{2017}\natexlab{a}.
\newblock \showarticletitle{Food Recommender Systems: Important Contributions,
  Challenges and Future Research Directions}.
\newblock \bibinfo{journal}{\emph{CoRR}}  \bibinfo{volume}{abs/1711.02760}
  (\bibinfo{year}{2017}).
\newblock
\showeprint[arxiv]{1711.02760}
\urldef\tempurl%
\url{http://arxiv.org/abs/1711.02760}
\showURL{%
\tempurl}


\bibitem[\protect\citeauthoryear{Trattner and Elsweiler}{Trattner and
  Elsweiler}{2017b}]%
        {Trattner:IHI:3038912:3052573}
\bibfield{author}{\bibinfo{person}{Christoph Trattner} {and}
  \bibinfo{person}{David Elsweiler}.} \bibinfo{year}{2017}\natexlab{b}.
\newblock \showarticletitle{Investigating the Healthiness of Internet-Sourced
  Recipes: Implications for Meal Planning and Recommender Systems}. In
  \bibinfo{booktitle}{\emph{Proceedings of the 26th International Conference on
  World Wide Web}} \emph{(\bibinfo{series}{WWW '17})}.
  \bibinfo{publisher}{International World Wide Web Conferences Steering
  Committee}, \bibinfo{address}{Republic and Canton of Geneva, Switzerland},
  \bibinfo{pages}{489--498}.
\newblock
\showISBNx{978-1-4503-4913-0}
\urldef\tempurl%
\url{https://doi.org/10.1145/3038912.3052573}
\showDOI{\tempurl}


\bibitem[\protect\citeauthoryear{Trattner, Kusmierczyk, and Nørvåg}{Trattner
  et~al\mbox{.}}{2019}]%
        {TRATTNER:2019654}
\bibfield{author}{\bibinfo{person}{Christoph Trattner}, \bibinfo{person}{Tomasz
  Kusmierczyk}, {and} \bibinfo{person}{Kjetil Nørvåg}.}
  \bibinfo{year}{2019}\natexlab{}.
\newblock \showarticletitle{Investigating and predicting online food recipe
  upload behavior}.
\newblock \bibinfo{journal}{\emph{Information Processing and Management}}
  \bibinfo{volume}{56}, \bibinfo{number}{3} (\bibinfo{year}{2019}),
  \bibinfo{pages}{654 -- 673}.
\newblock
\showISSN{0306-4573}
\urldef\tempurl%
\url{https://doi.org/10.1016/j.ipm.2018.10.016}
\showDOI{\tempurl}


\bibitem[\protect\citeauthoryear{Trattner, Moesslang, and Elsweiler}{Trattner
  et~al\mbox{.}}{2018}]%
        {Trattner2018}
\bibfield{author}{\bibinfo{person}{Christoph Trattner},
  \bibinfo{person}{Dominik Moesslang}, {and} \bibinfo{person}{David
  Elsweiler}.} \bibinfo{year}{2018}\natexlab{}.
\newblock \showarticletitle{On the predictability of the popularity of online
  recipes}.
\newblock \bibinfo{journal}{\emph{EPJ Data Science}} \bibinfo{volume}{7},
  \bibinfo{number}{1} (\bibinfo{date}{05 Jul} \bibinfo{year}{2018}),
  \bibinfo{pages}{20}.
\newblock
\showISSN{2193-1127}
\urldef\tempurl%
\url{https://doi.org/10.1140/epjds/s13688-018-0149-5}
\showDOI{\tempurl}


\bibitem[\protect\citeauthoryear{Trattner, Rokicki, and Herder}{Trattner
  et~al\mbox{.}}{2017}]%
        {Trattner:2017:RCI:3099023.3099072}
\bibfield{author}{\bibinfo{person}{Christoph Trattner}, \bibinfo{person}{Markus
  Rokicki}, {and} \bibinfo{person}{Eelco Herder}.}
  \bibinfo{year}{2017}\natexlab{}.
\newblock \showarticletitle{On the Relations Between Cooking Interests, Hobbies
  and Nutritional Values of Online Recipes: Implications for Health-Aware
  Recipe Recommender Systems}. In \bibinfo{booktitle}{\emph{Adjunct Publication
  of the 25th Conference on User Modeling, Adaptation and Personalization}}
  \emph{(\bibinfo{series}{UMAP '17})}. \bibinfo{publisher}{ACM},
  \bibinfo{address}{New York, NY, USA}, \bibinfo{pages}{59--64}.
\newblock
\showISBNx{978-1-4503-5067-9}
\urldef\tempurl%
\url{https://doi.org/10.1145/3099023.3099072}
\showDOI{\tempurl}


\bibitem[\protect\citeauthoryear{Ueta, Iwakami, and Ito}{Ueta
  et~al\mbox{.}}{2011}]%
        {Ueta:2011:RRS:2186633.2186642}
\bibfield{author}{\bibinfo{person}{Tsuguya Ueta}, \bibinfo{person}{Masashi
  Iwakami}, {and} \bibinfo{person}{Takayuki Ito}.}
  \bibinfo{year}{2011}\natexlab{}.
\newblock \showarticletitle{A Recipe Recommendation System Based on Automatic
  Nutrition Information Extraction}. In \bibinfo{booktitle}{\emph{Proceedings
  of the 5th International Conference on Knowledge Science, Engineering and
  Management}} \emph{(\bibinfo{series}{KSEM'11})}.
  \bibinfo{publisher}{Springer-Verlag}, \bibinfo{address}{Berlin, Heidelberg},
  \bibinfo{pages}{79--90}.
\newblock
\showISBNx{978-3-642-25974-6}
\urldef\tempurl%
\url{https://doi.org/10.1007/978-3-642-25975-3_8}
\showDOI{\tempurl}


\bibitem[\protect\citeauthoryear{van Pinxteren, Geleijnse, and Kamsteeg}{van
  Pinxteren et~al\mbox{.}}{2011}]%
        {vanPinxteren:2011:DRS:1943403.1943422}
\bibfield{author}{\bibinfo{person}{Youri van Pinxteren}, \bibinfo{person}{Gijs
  Geleijnse}, {and} \bibinfo{person}{Paul Kamsteeg}.}
  \bibinfo{year}{2011}\natexlab{}.
\newblock \showarticletitle{Deriving a Recipe Similarity Measure for
  Recommending Healthful Meals}. In \bibinfo{booktitle}{\emph{Proceedings of
  the 16th International Conference on Intelligent User Interfaces}}
  \emph{(\bibinfo{series}{IUI '11})}. \bibinfo{publisher}{ACM},
  \bibinfo{address}{New York, NY, USA}, \bibinfo{pages}{105--114}.
\newblock
\showISBNx{978-1-4503-0419-1}
\urldef\tempurl%
\url{https://doi.org/10.1145/1943403.1943422}
\showDOI{\tempurl}


\end{thebibliography}

\end{document}